\documentclass[twocolumn,english,aps,prb,twocolum,superscriptaddress,bibnotes,amsmath,amssymb,floatfix]{revtex4-1}
\usepackage[colorlinks=true,citecolor=blue,linkcolor=magenta]{hyperref}

\usepackage[markup=nocolor, authormarkupposition=left]{changes} 
\usepackage{soul}
\usepackage[utf8]{inputenc}
\usepackage[english]{babel}
\usepackage{amsmath,amsfonts,amssymb} 
\usepackage[T1]{fontenc}
\usepackage{url}
\usepackage{makecell}

\usepackage{amsmath}
\usepackage{amsfonts}
\usepackage{amssymb}

\usepackage{epstopdf}
\usepackage{graphicx}
\graphicspath{{./Figures/}}

\usepackage{changes}

\begin{document}
\title{Frequency-comb-linearized, widely tunable lasers for coherent ranging}

\author{Baoqi Shi}
\thanks{These authors contributed equally to this work.}
\affiliation{Department of Optics and Optical Engineering, University of Science and Technology of China, Hefei, Anhui 230026, China}
\affiliation{International Quantum Academy, Shenzhen 518048, China}

\author{Yi-Han Luo}
\thanks{These authors contributed equally to this work.}
\affiliation{International Quantum Academy, Shenzhen 518048, China}
\affiliation{Shenzhen Institute for Quantum Science and Engineering, Southern University of Science and Technology,
Shenzhen 518055, China}

\author{Wei Sun}
\affiliation{International Quantum Academy, Shenzhen 518048, China}

\author{Yue Hu}
\affiliation{International Quantum Academy, Shenzhen 518048, China}
\affiliation{Shenzhen Institute for Quantum Science and Engineering, Southern University of Science and Technology,
Shenzhen 518055, China}

\author{Jinbao Long}
\affiliation{International Quantum Academy, Shenzhen 518048, China}

\author{Xue Bai}
\affiliation{International Quantum Academy, Shenzhen 518048, China}

\author{Anting Wang}
\affiliation{Department of Optics and Optical Engineering, University of Science and Technology of China, Hefei, Anhui 230026, China}

\author{Junqiu Liu}
\email[]{liujq@iqasz.cn}
\affiliation{International Quantum Academy, Shenzhen 518048, China}
\affiliation{Hefei National Laboratory, University of Science and Technology of China, Hefei 230088, China}

\maketitle

\noindent\textbf{
\noindent Tunable lasers, with the ability to continuously adjust their emission wavelengths, have found widespread applications across various fields such as biomedical imaging, coherent ranging, optical communications and spectroscopy. 
In these applications, a wide chirp range is advantageous for large spectral coverage and high frequency resolution.
Besides, the frequency accuracy and precision also depend critically on the chirp linearity of the laser.
While extensive efforts have been made on the development of many kinds of frequency-agile, widely tunable, narrow-linewidth lasers, wideband yet precise methods to characterize and to linearize laser chirp dynamics are also demanded.
Here we present an approach to characterize laser chirp dynamics using an optical frequency comb.
The instantaneous laser frequency is tracked over terahertz bandwidth with 1 MHz interval.
Using this approach we calibrate the chirp performance of twelve tunable lasers from Toptica, Santec, New Focus, EXFO and NKT that are commonly used in fiber optics and integrated photonics.
In addition, with acquired knowledge on laser chirp dynamics, we demonstrate a simple frequency-linearization scheme that enables coherent ranging without any optical or electronic linearization units. 
Our approach not only presents a novel wideband, high-resolution laser spectroscopy, but is also critical for sensing applications with ever-increasing requirements on performance.  
}

\begin{figure*}[t!]
\centering
\includegraphics{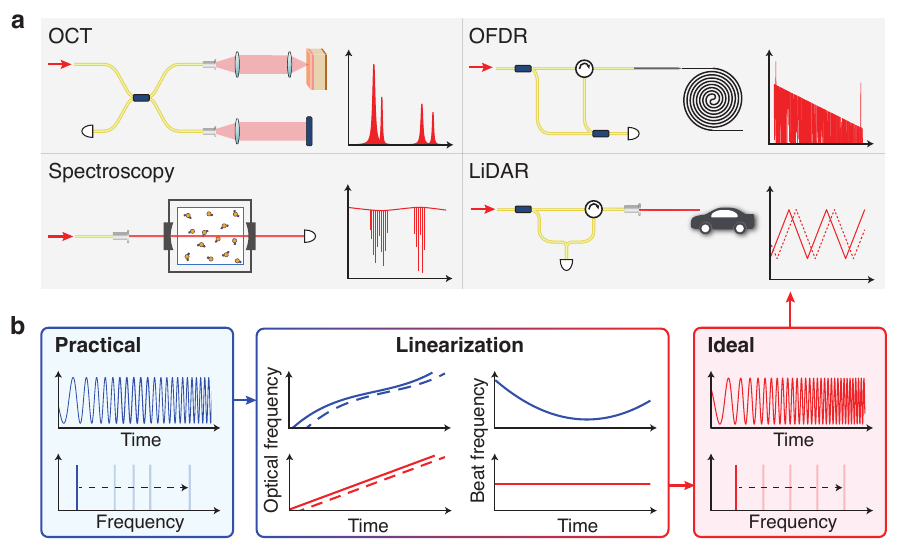}
\caption{
\textbf{Principle and applications of widely tunable lasers}.
\textbf{a}. 
Applications that require linearly chirping lasers.
OCT, optical coherence tomography. 
OFDR, optical frequency-domain reflectometry. 
LiDAR, light detection and ranging.
\textbf{b}.
Principle of laser chirp linearization. 
An ideal laser chirps at a constant rate.
However, in reality the actual chirp rate varies.
By beating the laser with its delayed part, the chirp nonlinearity in the optical domain is revealed in the RF domain. 
}
\label{Fig:1}
\end{figure*}
\begin{figure*}[t!]
\centering
\includegraphics{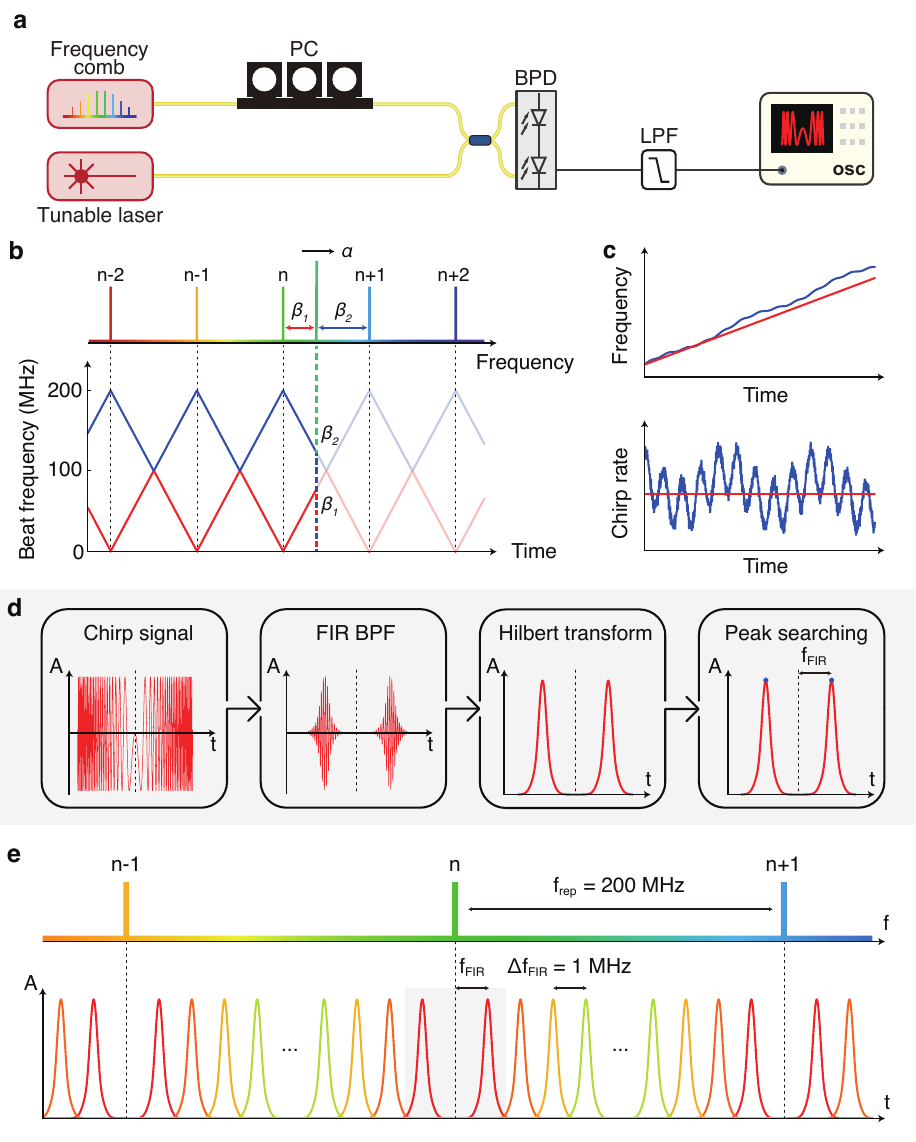}
\caption{
\textbf{Schematic and experimental setup of laser chirp characterization}.
\textbf{a}. 
Experimental setup. 
BPF, band-pass filter. 
PC, polarization controller.
BPD, balanced photodetector. 
LPF, low-pass filter. 
OSC, oscilloscope. 
\textbf{b}. 
Illustration of the laser frequency beating with the OFC during laser chirping at a rate of $\alpha$. 
The time trace of the two beat frequencies $\beta_1$ and $\beta_2$ ($\beta_2>\beta_1$) are shown. 
\textbf{c}. 
Upper panel shows the instantaneous frequency of the chirping laser in the ideal (red) and actual (blue) cases.
Lower panel shows the corresponding instantaneous chirp rate $\alpha$.
\textbf{d}. 
Flowcharts of the algorithm based on finite impulse response (FIR) band-pass filters, to extract the instantaneous laser frequency as well as the chirp rate. 
The dashed lines mark the region where the laser frequency scans across a comb line.
\textbf{e}. 
The FIR filter's center frequency $f_\mathrm{FIR}$ is digitally set, and the instantaneous laser frequency is calculated over 1 MHz interval.
}
\label{Fig:2}
\vspace{3cm}
\end{figure*}
\noindent \textbf{Introduction}. 
Tunable lasers, which have the ability to dynamically adjust their emission wavelength, have found widespread applications in a variety of industries and scientific fields.
For example, as shown in Fig. \ref{Fig:1}a, in bio-medical imaging, optical coherence tomography (OCT) \cite{Cense:04, Grulkowski:12} employs a widely tunable laser that provides a non-invasive and non-contact imaging modality for high-resolution two-dimensional cross-sectional and three-dimensional volumetric imaging of tissue structures.
In optical communication systems, optical frequency-domain reflectometry (OFDR) \cite{Lexe:97, Soller:05} uses tunable lasers for broadband loss and dispersion characterization.  
In laser spectroscopy, tunable diode laser spectrometers (TDLAS) \cite{LiuX:22, Liu:16} allow compound analysis of gases, as well as their conditions such as concentration, pressure, temperature and velocity.
For light detection and ranging (LiDAR), frequency-modulated continuous-wave (FMCW) LiDAR \cite{Roos:09, Baumann:13, Lihachev:22} uses a tunable laser to measure the distance (based on the time of flight) and speed (based on the Doppler effect) of a moving object.

In all these applications, as well as others, the laser chirp range determines the measurement resolution and spectral bandwidth. 
However, due to the nonlinearity of laser gain and dynamics, tunable lasers typically exhibit chirping nonlinearity that is particularly deteriorated for a wide chirp range, as illustrated in Fig. \ref{Fig:1}b.
As a result, the laser chirp rate deviates from the set value, which compromises the frequency resolution, precision and accuracy in applications.
Therefore, widely tunable lasers require careful characterization and linearization for demanding applications.

To trace the laser chirp rate $\alpha(t)$, a fiber Mach-Zehnder interferometer (MZI) is commonly used to calibrate the instantaneous frequency, e.g. in data processing \cite{Glombitza:93, Ahn:05}, active linearization \cite{Roos:09}, and laser drive signal optimization\cite{ZhangX:19}.
However, for a wide chirp range, the fiber dispersion seriously compromises the calibration precision of fiber MZIs and cause phase-instability issues, leading to insufficient accuracy for linearization.
In comparison, optical frequency combs (OFC) with equidistant grids of frequency lines can essentially resolve these issues \cite{DelHaye:09, Giorgetta:10, Baumann:13, Twayana:21}.
Here we demonstrated an approach to characterize the chirp dynamics of widely tunable lasers. 
We apply this method on twelve lasers from Toptica, Santec, New Focus, EXFO and NKT that are commonly used in fiber optics and integrated photonics. 
Using an OFC and digital band-pass filters, the instantaneous laser frequency is tracked with every $1$ MHz interval over terahertz bandwidth. 
This allows the linearization of the pre-calibrated chirp rate $\alpha(t)$, particularly useful for coherent LiDAR.

\section{Principle and setup}

To trace the laser chirp rate $\alpha(t)$, we use a commercial, fully stabilized, fiber-based OFC (Quantum CTek) as a frequency ``ruler'' \cite{Udem:02, Cundiff:03, Diddams:20} to trace the instantaneous laser frequency during chirping. 
In our case, the comb repetition rate $f_\mathrm{rep}=200$ MHz and its carrier envelope offset frequency $f_\mathrm{ceo}$ are actively locked to a rubidium atomic clock, thus the OFC is fully stabilized. 
Consequently, in the frequency domain, the $n$-th comb line's frequency is unambiguously determined as $f_n=f_\mathrm{ceo}+n\cdot f_\mathrm{rep} (n \in \mathbb{N})$.

\begin{figure*}[t!]
\centering
\includegraphics{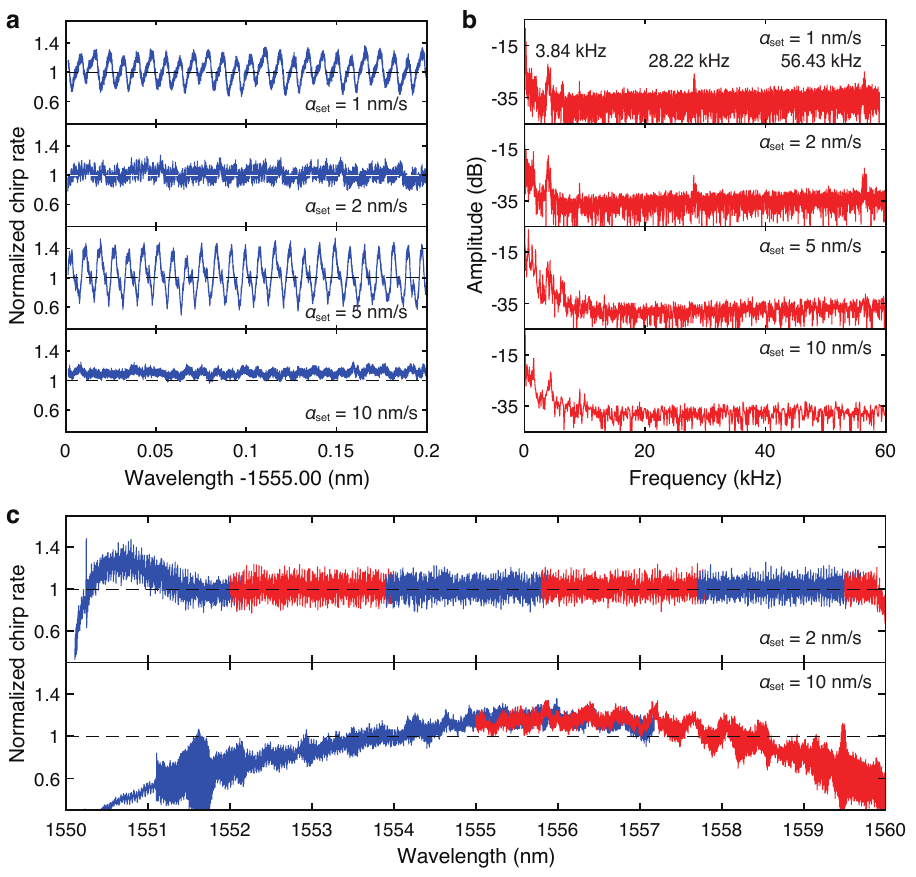}
\caption{
\textbf{Characterization of Toptica CTL laser's chirp dynamics}.
\textbf{a}.
Normalized chirp rate $\alpha(\lambda)/\alpha_\mathrm{set}$ with different set values $\alpha_\mathrm{set}$ over 0.2 nm wavelength range. 
\textbf{b}. 
Frequency spectrum of the normalized chirp rate $\alpha(t)/\alpha_\mathrm{set}$.
The frequency values of prominent peaks are marked. 
\textbf{c}. 
Normalized chirp rate $\alpha(\lambda)/\alpha_\mathrm{set}$ with different set values $\alpha_\mathrm{set}$ over 10 nm wavelength range.
}
\label{Fig:3}
\end{figure*}

When the laser chirps, it beats against tens of thousands of comb lines with precisely known frequency values, as shown in Fig. \ref{Fig:2}a.
Beat signals of frequencies $|f_n-f_l(t)|$ are generated, where $f_l(t)$ is the instantaneous frequency of the chirping laser at time $t$. 
As the frequency comb is fully stabilized, for a mode-hop-free chirping laser, the dynamics and linearity of $f_l(t)$ are projected onto the beat signals. 
The simplest beat signals to analyze are the two generated by the chirping laser with its two neighbouring comb lines, i.e. $|f_n-f_l(t)|\leq f_\mathrm{rep}=200$~MHz.
Experimentally, the two beat signals are recorded by an oscilloscope after passing through a low-pass filter (LPF, Mini-Circuits, BLP-150+), whose 3-dB cutoff frequency is 155 MHz in our case. 
We denote the frequencies of the two beat signals as $\beta_1$ and $\beta_2$ ($\beta_1+\beta_2=f_\mathrm{rep}=200$ MHz and $\beta_1<\beta_2$). 
As shown in Fig. \ref{Fig:2}b, for continuous linear chirp, when the laser frequency passes through the comb lines sequentially, $\beta_1$ and $\beta_2$ vary in a sawtooth waveform. 
The time-domain waveform near $\beta_1=0$ is shown in the leftmost panel of Fig. \ref{Fig:2}d, corresponding to the situation when the laser chirps across a comb line.

Here, we only concentrate on the analysis of $\beta_1$, since $\beta_1+\beta_2=200$ MHz.
The dynamics and linearity of $\beta_1(t)$ as a function of time are identical to those of the laser frequency $f_l(t)$. 
We process the $\beta_1(t)$ signal and extract the instantaneous frequency value as illustrated in Fig. \ref{Fig:2}d. 
A finite impulse response (FIR) band-pass filter with a specific pass-band center frequency $f_\text{FIR}$ is applied to the recorded $\beta_1(t)$ signal. 
The FIR filter only transmits the temporal segments when $\beta_1(t)$ is around $f_\text{FIR}$, as illustrated in the left middle panel of Fig. \ref{Fig:2}d. 
Then Hilbert transform \cite{Marple:99} is applied to obtain the pulse's envelope.
Via peak searching, the exact time of the pulses' centers is extracted and recorded, when $\beta_1(t)=f_\text{FIR}$. 
Experimentally, we repeatedly apply FIR filters whose $f_\text{FIR}$ values are digitally set from 3 MHz to 97 MHz with an interval of $\Delta f_\text{FIR}=1$ MHz, as shown in Fig. \ref{Fig:2}e. 
In this way, during continuous laser chirp over a wide bandwidth, the $\beta_1(t)$ time trace is frequency-calibrated and recorded with 1 MHz interval. 

Next we unwrap the sawtooth-like $\beta_1(t)$ trace. 
We select pulses corresponding to $f_\text{FIR} = 3$ MHz as markers. 
As shown in the bottom panel of Fig. \ref{Fig:2}e, when the laser chirps over a pair of comb lines (dashed lines), i.e. $f_\text{rep}$ distance, two markers are created by the $f_\text{FIR}=3$ MHz filters.
With the known comb spacing $f_\text{rep} = 200$ MHz, $\beta_1(t)$ can then be unwrapped to an increasing function of time $t$, which is $\Delta f(t) = f_l(t) - f_l(t_0)$, where $f_l(t_0)$ is the starting laser frequency at time $t_0$. 
As shown in Fig. \ref{Fig:2}c, the instantaneous chirp rate - estimated as the average rate within $\Delta f_\text{FIR}=1$ MHz calibration interval - is calculated as $\alpha(t)=\mathrm{d}\Delta f(t)/\mathrm{d}t$, and is later used for chirp linearization.

\section{Results}

Now we use this frequency-comb-calibrated method to characterize totally ten widely tunable, mode-hop-free, external-cavity diode lasers (ECDL), including \textbf{three Toptica CTL} lasers, \textbf{four Santec} lasers, \textbf{two New Focus} lasers, and \textbf{one EXFO} laser. 
These lasers are all operated in the telecommunication band around 1550 nm, and are extensively used in fiber optics and integrated photonics. 
For example, Toptica CTL and New Focus lasers are widely used in the generation of microresonator-based dissipative Kerr soliton frequency combs \cite{Herr:14, Guo:16, Liu:20}. 
Santec and EXFO lasers are widely used in wideband characterization of waveguide dispersion \cite{Liu:16, Twayana:21, Luo:23}. 
In addition, \textbf{two NKT} fiber lasers are characterized, although their frequency tuning ranges are significantly narrower.

An ECDL typically has two tuning modes, i.e. the \textit{wide} and \textit{fine} tuning modes.
In the wide tuning mode, the external cavity length that determines the laser frequency is controlled by a stepper motor, enabling a frequency tuning range exceeding ten terahertz.
In the fine tuning mode, the external cavity length is controlled by a piezo under an external voltage that determines the laser frequency.
In this mode, the laser can only be tuned by tens of gigahertz.
Triangular and sinusoidal voltage signals are often used to drive the piezo, and to determine the laser chirp range $B$ and modulation frequency $f_\mathrm{mod}$.
The average chirp rate is $\overline{\alpha_\mathrm{set}}=2B f_\mathrm{mod}$.

\begin{table*}[t!]
\centering
\caption{Comparison of laser chirp dynamics of different lasers with with different conditions.}
\label{tab:average}
\begin{tabular}{|c|c|c|c|c|c|c|c|}
\hline
Laser brand & Model & Tuning mode & Optimal $\alpha_\mathrm{set}$ & $f_\text{mod}$ & Chirp range & RMSE & $\delta_\mathrm{L}$ \\
\hline
Toptica & CTL 1550 & Wide, single & 2 nm/s (0.24 THz/s) & - & 10 nm (1.2 THz) & 7.5\% (1.1\%) & 0.2\% (0.0\%)\\
Santec & TSL-570-A & Wide, single & 100 nm/s (12 THz/s) & - & 10 nm  (1.2 THz) & 6.8\% (1.6\%) & 0.4\% (0.2\%)\\
New Focus & TLB-6700 & Wide, single & 2 nm/s (0.24 THz/s) & - & 10 nm  (1.2 THz) & 12.9\% (1.9\%) & 0.3\% (0.2\%)\\
EXFO & T500S & Wide, single & 100 nm/s (12 THz/s) & - & 10 nm  (1.2 THz) & 4.8\% & 0.2\%\\
\hline
Toptica & CTL 1550 & Fine, triangular & 0.4 nm/s (50 GHz/s) & 10 Hz & 80 pm (10 GHz) & 15.8\% (3.8\%) &  0.8\% (0.0\%)\\
Santec & TSL-570-A & Fine, triangular & 2 nm/s (250 GHz/s) & 50 Hz & 80 pm (10 GHz) & 21.3\% (0.7\%) &  2.8\% (0.1\%)\\
New Focus & TLB-6700 & Fine, triangular & 2 nm/s (250 GHz/s) & 50 Hz & 80 pm (10 GHz) & 18.2\% (11.8\%) &  0.4\% (0.1\%)\\
NKT & E15 & Fine, triangular & 64 pm/s (8 GHz/s) & 2 Hz & 64 pm (8 GHz) & 5.9\% (1.3\%) &  0.5\% (0.3\%)\\
\hline
Toptica & CTL 1550 & Fine, sinusoidal & 2 nm/s (250 GHz/s) & 50 Hz & 80 pm (10 GHz) & 15.8\% (3.2\%) & -\\
Santec & TSL-570-A & Fine, sinusoidal & 2 nm/s (250 GHz/s) & 50 Hz & 80 pm (10 GHz) & 26.1\% (1.4\%) & -\\
New Focus & TLB-6700 & Fine, sinusoidal & 4 nm/s (500 GHz/s) & 100 Hz & 80 pm (10 GHz) & 8.7\% (3.8\%) & -\\
NKT & E15 & Fine, sinusoidal & 64 pm/s (8 GHz/s) & 2 Hz & 64 pm (8 GHz) & 7.1\% (0.5\%) & -\\
\hline
\end{tabular}
\end{table*}

First, we characterize these lasers in the wide tuning mode.
We take one Toptica CTL laser as the first example. 
Experimentally, the laser is configured to a single run that chirps from 1550 to 1560 nm.
On the laser panel, we manually set the chirp rate $\alpha_\mathrm{set}$ to 1, 2, 5, and 10 nm/s.
The oscilloscope's sample resolution is set to 155 Hz for different $\alpha_\mathrm{set}$.
The instantaneous chirp rate $\alpha(t)$ within 1 MHz interval is tracked, allowing retrieval of the instantaneous laser frequency $f_l(t)$ as well as the wavelength $\lambda(t)$.
In this manner, $\alpha(t)$ can be convert to $\alpha(\lambda)$ for a better comparison of cases of different $\alpha_\mathrm{set}$.

Figure \ref{Fig:3} shows representative $\alpha(\lambda)$ traces with different $\alpha_\mathrm{set}$ within 0.2 nm wavelength range (1555 to 1555.2 nm) during chirping.  
To facilitate comparison, we normalize the measured $\alpha(\lambda)$ to the set value $\alpha_\mathrm{set}$ on the laser panel, i.e. the normalized chirp rate is $\alpha(\lambda)/\alpha_\mathrm{set}$. 
Here $\alpha(\lambda)/\alpha_\mathrm{set}=1$ corresponds to perfectly linear chirp at a constant rate.
Figure \ref{Fig:3}a shows that, the experimentally measured $\alpha(\lambda)/\alpha_\mathrm{set}$ fluctuates and exhibits nonlinearity.  
Particularly, when $\alpha_\mathrm{set}=1$ and 5 nm/s, prominent frequency oscillation is observed.
Whereas for $\alpha_\mathrm{set}=2$ and 10 nm/s, this oscillation is inhibited.
The oscillation pattern of $\alpha(\lambda)/\alpha_\mathrm{set}$ is repeated every 1.6 pm, probably due to the laser's intrinsic configuration or regulation.
Besides, mechanical modes of the external laser cavity might also cause chirp rate jitter.
Figure \ref{Fig:3}b shows $\alpha(t)/\alpha_\mathrm{set}$ spectra in the frequency domain. 
A component at 3.86 kHz Fourier offset frequency - independent of $\alpha_\mathrm{set}$ - is revealed, likely linking to the eigen-frequency of the cavity's fundamental mechanical mode.
Higher-order mechanical modes can also be observed in Fig. \ref{Fig:3}b at 28.22 and 56.43 kHz frequencies when the laser chirps more slowly. 

Since Fig. \ref{Fig:3}a shows the optimal laser chirping performance at $\alpha_\mathrm{set}=2$ and 10 nm/s, we further investigate $\alpha(\lambda)/\alpha_\mathrm{set}$ over the entire 10 nm bandwidth.
Figure \ref{Fig:3}c shows the laser's chirp rate drift with $\alpha_\mathrm{set}=2$ and $10$ nm/s. 
Limited by the oscilloscope's memory depth (400 Mpts) and sampling rate (400 MSa/s), we can only acquire data in 1 second.
We set the laser to repeatedly chirp from 1550 to 1560 nm, and characterize one segment of the chirp trace each time.
Multiple segments displayed with different colors are stitched to form a complete trace covering the full 10 nm range.
It becomes apparent that, although Fig. \ref{Fig:3}b bottom shows weak chirp rate jitter with $\alpha_\mathrm{set}=10$ nm/s, the overall chirp rate drifts considerably.
With $\alpha_\mathrm{set}=2$ nm/s, the laser completes its acceleration at $1550.5$ nm and reaches stability at 1552.0 nm, until it begins to decelerate at 1559.5 nm. 
The laser maintains a good linearity in the center 7.5 nm range.
We therefore conclude that, in the wide tuning mode, the optimal chirp rate of Toptica CTL is $\alpha_\mathrm{set} = 2$ nm/s.

\begin{figure*}[t!]
\centering
\includegraphics{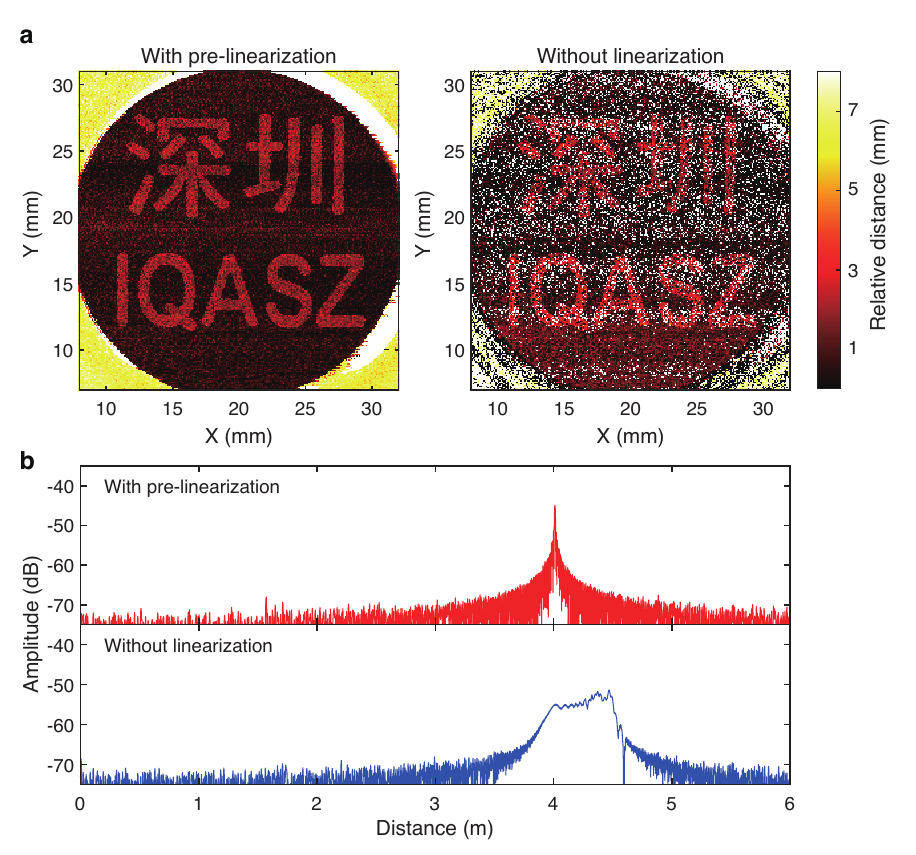}
\caption{
\textbf{Coherent LiDAR experiment}.
\textbf{a}.
Maps of measured relative distance of an engraved surface with pre-linearization and without linearization.
A tilt angle of $11.6^{\circ}$ of the sample is measured and further subtracted.
\textbf{b}.
Space spectra of the ranging profiles obtained with pre-linearization and without linearization. 
Fast Fourier transform is applied on the ranging profile of each case to retrieve the spectrum. 
The peaks correspond to the distance $d=4$ m of the surface to the laser.
}
\label{Fig:4}
\end{figure*}

We further characterized another two Toptica CTL lasers, four Santec lasers, two New Focus lasers and one EXFO laser in their wide tuning modes.
The laser chirping performance is investigated within 1550 to 1560 nm bandwidth, and $\alpha_\mathrm{set}$ is set from 1 to 200 nm/s.
Table \ref{tab:average} presents a summary of different lasers' chirp dynamics with their respective optimal $\alpha_\mathrm{set}$. 
We refer the readers to Supplementary Materials for detailed characterization results of each laser, which can be useful for readers currently using these lasers.
Overall, in the wide tuning mode, the Santec and EXFO lasers work best with $\alpha_\mathrm{set}=100$ nm/s, and the Toptica CTL and New Focus lasers work best with $\alpha_\mathrm{set}=2$ nm/s.

To show the deviation of laser chirp rate from $\alpha(t)/\alpha_\mathrm{set}=1$, we calculate the root mean square error (RMSE) as
\begin{equation}
\mathrm{RMSE}=\sqrt{\frac{\sum_{i=1}^T (\alpha_i/\alpha_\mathrm{set}-1)^2}{T}},
\label{Eq.RMSE}
\end{equation}
where $\alpha_i$ is the sample of chirp rate and $T$ is sample number.
Besides, the wavelength linearity can be estimated by nonlinearity error \cite{Giusca:12} $\delta_\mathrm{L} = \Delta \lambda_\mathrm{max}/\lambda_\mathrm{f.s.}$, where $\Delta \lambda_\mathrm{max}$ is the maximum deviation of the measured wavelength from its linearly fitted value, and $\lambda_\mathrm{f.s.}$ is the full-scale wavelength range.
This indicator mainly considers whether the laser chirp is linear, but not whether the average chirp rate $\overline{\alpha_\mathrm{set}}$ matches the set value $\alpha_\mathrm{set}$.
Particularly, Table \ref{tab:average} shows that the EXFO laser features the smallest RMSE and $\delta_\mathrm{L}$ with $\alpha_\mathrm{set}=100$ nm/s. 

Next, we characterized eleven lasers in the fine tuning modes, including three Toptica CTL lasers, four Santec lasers, two New Focus lasers, and two NKT lasers. 
We note that the EXFO laser does not feature the fine tuning mode. 
These lasers are frequency-modulated at 193.3 THz optical frequency and with excursion range of 10 GHz (except that 8 GHz for the NKT lasers). 
The drive voltage signal to the piezo is either triangular or sinusoidal.
We set the modulation frequency $f_\mathrm{mod}$ to 2, 10, 50 and 100 Hz.
For normalization, the set chirp rate $\alpha_\mathrm{set}(t)$ is calculated according to the drive signal.
The optimal $f_\mathrm{mod}$ for different lasers with different drive manners are listed in Table \ref{tab:average} (see Supplementary Materials for detailed characterization results of each laser).
For the cases with sinusoidal drive, $\alpha_\mathrm{set}$ represents the average set chirp rate $\overline{\alpha_\mathrm{set}}$.
For the Toptica CTL and New Focus lasers, different drives have different effects, as the triangular drive may cause resonance in the piezo at certain frequencies (see Supplementary Materials).
Therefore, these lasers behave better with a sinusoidal drive.

\noindent \textbf{Light detection and ranging}. 
Based on our frequency-comb-calibrated laser chirping dynamics, we further demonstrate a linearization method for widely chirping laser for FMCW LiDAR. 
LiDAR can quickly and accurately map the environment, provide insights into the structure and composition, and allow operators to make informed decisions on the protection of natural resources \cite{Lefsky:02,Simard:11} and environment \cite{Fernald:84,Vaughan:09} or manage agriculture \cite{Mulla:13}. 
For autonomous vehicles, LiDAR enables safe and reliable navigation by providing high-resolution, real-time 3D mapping of the environment \cite{Chen:17,Yue:18}.
The technology has also revolutionized archaeology, allowing non-destructive imaging of hidden features and artifacts \cite{Chase:11,Chase:12,EvansD:13}. 

The principle of LiDAR is to project an optical signal onto a moving object. 
The reflected or scattered signal is received and processed to determine the object's distance from the laser \cite{Amann:01}.
The light received by the photodetector experiences a time delay $\tau$ from its emission time.
Thus the object's distance is calculated as $d=c\tau/2$, where $c$ is the speed of light. 
For FMCW LiDAR, the time delay $\tau$ is obtained in a coherent way \cite{Roos:09,Baumann:13,Lihachev:22}.
The frequency-modulated laser is split into two paths as shown in the LiDAR panel of Fig. \ref{Fig:1}a.
In one path, the laser passes through a collimator, enters into free space, and is reflected by the object. 
The reflected laser is then combined with the other reference path, and the beat signal between the two paths is recorded by a photodetector. 
If the laser frequency is modulated linearly at a constant chirp rate $\alpha$, the time delay $\tau$ creates a beat signal of frequency $\Delta f=\alpha \tau$.
Thus, the photodetected beat signal is
\begin{equation}
\begin{aligned}
V(t)&\propto\cos(2\pi\Delta f t)\\
&=\cos\left(2\pi\alpha\frac{2d}{c} t\right).
\end{aligned}
\label{Eq.Vt}
\end{equation}
Equation \ref{Eq.Vt} indicates that the distance $d$ can be obtained with fast Fourier transformation on $t$.  

The ranging resolution \cite{Glombitza:93} $\delta d$ of FMCW LiDAR is limited by the chirp range $B$, as $\delta d=c/2B$.
High resolution, i.e. small $\delta d$, requires a large $B$ of the chirping laser. 
However, chirp nonlinearity causes a varying $\alpha$ over time.
Consequently, the beat frequency $\Delta f$ varies as shown in Fig.~\ref{Fig:1}b.
To extract the precise value of $d$ from $V(t)$, an accurate trace of $\alpha(t)$ is mandatory, necessitating the calibration of the instantaneous laser frequency during chirping.
The chirp linearization is performed by re-scaling the beat signal's time axis by $t'=\alpha(t) t$ to
\begin{equation}
V(t')=\cos\left(2\pi\frac{2d}{c} t'\right).
\label{Eq.Vt'}
\end{equation}

Here, using our frequency-comb-calibration method, the characterized laser chirp rate $\alpha(t)$ in return allows chirp rate pre-linearization. 
Again we use the Toptica CTL laser as the example.
The laser has optimal performance of chirp linearity when driven by a 50 Hz sinusoidal wave.
The chirp range is set to 35 GHz in the fine tuning mode, and $\alpha(t)$ is characterized (see Supplementary Materials). 
The same characterization of $\alpha(t)$ is performed three times, and the averaged result is used as the pre-linearized chirp rate $\alpha_\mathrm{c}(t)$.
By re-scaling the time axis in Eq. \ref{Eq.Vt} by $t'=\alpha_\mathrm{c}(t) t$, the laser is pre-linearized and the range profile can be extracted from Eq. \ref{Eq.Vt'}. 
The photodetected signal is re-scaled with $\alpha_\mathrm{c}(t)$.
Zero-padding \cite{Shinpaugh:92} is also implemented to reduce resolution ambiguities. 

Finally, we proceed a LiDAR experiment without any extra linearization unit.
Experimentally, a stainless steel cylinder with an engraved text of 2.8 mm depth on its surface is imaged using our LiDAR setup. 
The stainless steel cylinder is placed on a 2D translation stage 4 meter away from the collimator. 
As the translation stage moves, the distance between the cylinder surface and the collimator is continuously measured to produce a 2D ranging map. 

To highlight the effect of linearization, we measure the surface profile using two methods, i.e. with pre-linearization and without any linearization.
We emphasize that, the former case only requires a characterized $\alpha_\mathrm{c}(t)$ as prior knowledge for signal processing, and no OFC nor real-time frequency calibration is used in the ranging experiment. 
To retrieve the ranging profile, for pre-linearization, the beat signal $V(t')$ of Eq. \ref{Eq.Vt'} is rewritten as $V(l')=cos(2\pi dl')$, where $l'=ct'/2$.
By fast Fourier transform on $l'$, the space spectrum is obtained, as shown in the top panel of Fig. \ref{Fig:4}b.
The peak corresponds to the distance $d=4$ m of the surface to the laser.
For the case without linearization, the beat signal $V(t)$ of Eq. \ref{Eq.Vt} is rewritten as $V(l)=cos(2\pi dl)$, where $l=ct/2\alpha_\text{set}$ and $\alpha_\text{set}$ is the set chirp rate.
The space spectrum obtained by fast Fourier transform on $l$ is shown in the bottom panel of Fig. \ref{Fig:4}b.
Without linearization, the peak is broadened due to ranging imprecision, leading to ambiguous determination of distance.
Figure \ref{Fig:4}a shows the measured relative distance map with pre-linearization (left) and without any linearization (right). 
The long-term stability of pre-linearization is verified by a precision test (see Supplementary Materials).

\section{Conclusion}

In conclusion, we have demonstrated an approach to characterize chirp dynamics of widely tunable lasers based on an OFC.
Using this method we have characterized twelve lasers including three Toptica CTL lasers, four Santec lasers, two New Focus lasers, two NKT fiber lasers and one EXFO laser.
By comparing the laser's chirp linearity with different settings, the optimal operation condition of each laser is found. 
For example, the optimal chirp rate of Toptica CTL, New Focus, Santec, and EXFO lasers are $\alpha_\mathrm{set}=2$, 2, 100 and 100 nm/s, respectively.
Operating the laser with its optimal setting and pre-linearizing laser chirp, we successfully apply the laser for coherent LiDAR without any extra linearization units. 
Field-programmable gate arrays (FPGA) to implement multiple FIR filters can be further added to increase data processing speed and reduce the amount of data to be stored by real-time data processing.
Our method of characterizing and pre-linearizing laser chirp dynamics is proved to be a critical diagnostic method for applications such as OCT, OFDR, and TDLAS.
\medskip
\begin{footnotesize}

\noindent \textbf{Funding Information}: 
J. Liu acknowledges support from the National Natural Science Foundation of China (Grant No.12261131503), Shenzhen-Hong Kong Cooperation Zone for Technology and Innovation (HZQB-KCZYB2020050), and from the Guangdong Provincial Key Laboratory (2019B121203002).
Y.-H L. acknowledges support from the China Postdoctoral Science Foundation (Grant No. 2022M721482). 

\noindent \textbf{Author contributions}: 
B. S., W. S., Y.-H. L. and J. Long built the experimental setup, with the assistance from X. B. and A. W.. 
B. S. performed the laser characterization. 
B. S. and Y. H. and Y.-H. L. performed the LiDAR experiments. 
B. S., W. S., Y.-H. L. and J. Liu analysed the data and prepared the manuscript with input from others. 
J. Liu supervised the project.  

\noindent \textbf{Conflict of interest}:
B. S., Y.-H. L., W. S., X. B. and J. Liu filed a patent application for the tunable laser characterization and linearization method.  
Others declare no conflicts of interest. 

\noindent \textbf{Data Availability Statement}: 
The code and data used to produce the plots within this work will be released on the repository \texttt{Zenodo} upon publication of this preprint.

\end{footnotesize}
%
\end{document}


\title{Supplementary Materials for:\\ 
Frequency-comb-linearized, widely tunable lasers for coherent ranging}

\author{Baoqi Shi}
\thanks{These authors contributed equally to this work.}
\affiliation{Department of Optics and Optical Engineering, University of Science and Technology of China, Hefei, Anhui 230026, China}
\affiliation{International Quantum Academy, Shenzhen 518048, China}

\author{Yi-Han Luo}
\thanks{These authors contributed equally to this work.}
\affiliation{International Quantum Academy, Shenzhen 518048, China}
\affiliation{Shenzhen Institute for Quantum Science and Engineering, Southern University of Science and Technology,
Shenzhen 518055, China}

\author{Wei Sun}
\affiliation{International Quantum Academy, Shenzhen 518048, China}

\author{Yue Hu}
\affiliation{International Quantum Academy, Shenzhen 518048, China}
\affiliation{Shenzhen Institute for Quantum Science and Engineering, Southern University of Science and Technology,
Shenzhen 518055, China}

\author{Jinbao Long}
\affiliation{International Quantum Academy, Shenzhen 518048, China}

\author{Xue Bai}
\affiliation{International Quantum Academy, Shenzhen 518048, China}

\author{Anting Wang}
\affiliation{Department of Optics and Optical Engineering, University of Science and Technology of China, Hefei, Anhui 230026, China}

\author{Junqiu Liu}
\email[]{liujq@iqasz.cn}
\affiliation{International Quantum Academy, Shenzhen 518048, China}
\affiliation{Hefei National Laboratory, University of Science and Technology of China, Hefei 230088, China}

\maketitle

\section{Laser chirp dynamics for wide tuning mode}
\begin{figure*}[t!]
\renewcommand{\figurename}{Supplementary Figure}
\centering
\includegraphics{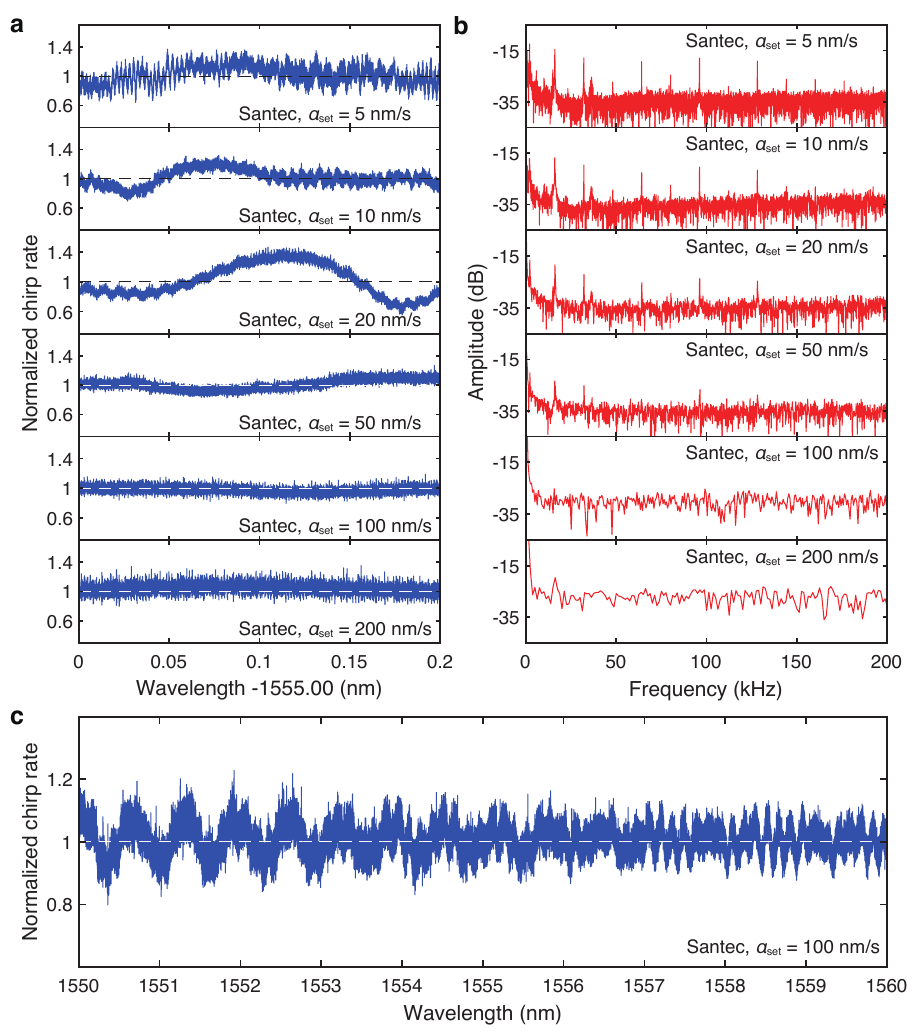}
\caption{
\textbf{Characterization of chirp dynamics of a Santec laser in wide tuning mode}.
\textbf{a}.
Normalized chirp rate $\alpha(\lambda)/\alpha_\mathrm{set}$ with different set values $\alpha_\mathrm{set}$ over 0.2 nm wavelength range.
\textbf{b}. 
Frequency spectrum of the chirp rate $\alpha(t)/\alpha_\mathrm{set}$.
\textbf{c}. 
Normalized chirp rate $\alpha(\lambda)/\alpha_\mathrm{set}$ with $\alpha_\mathrm{set} = 100$ nm/s over 10 nm wavelength range.
}
\label{SIfig:Santec_wide}
\end{figure*}

\begin{figure*}[t!]
\renewcommand{\figurename}{Supplementary Figure}
\centering
\includegraphics{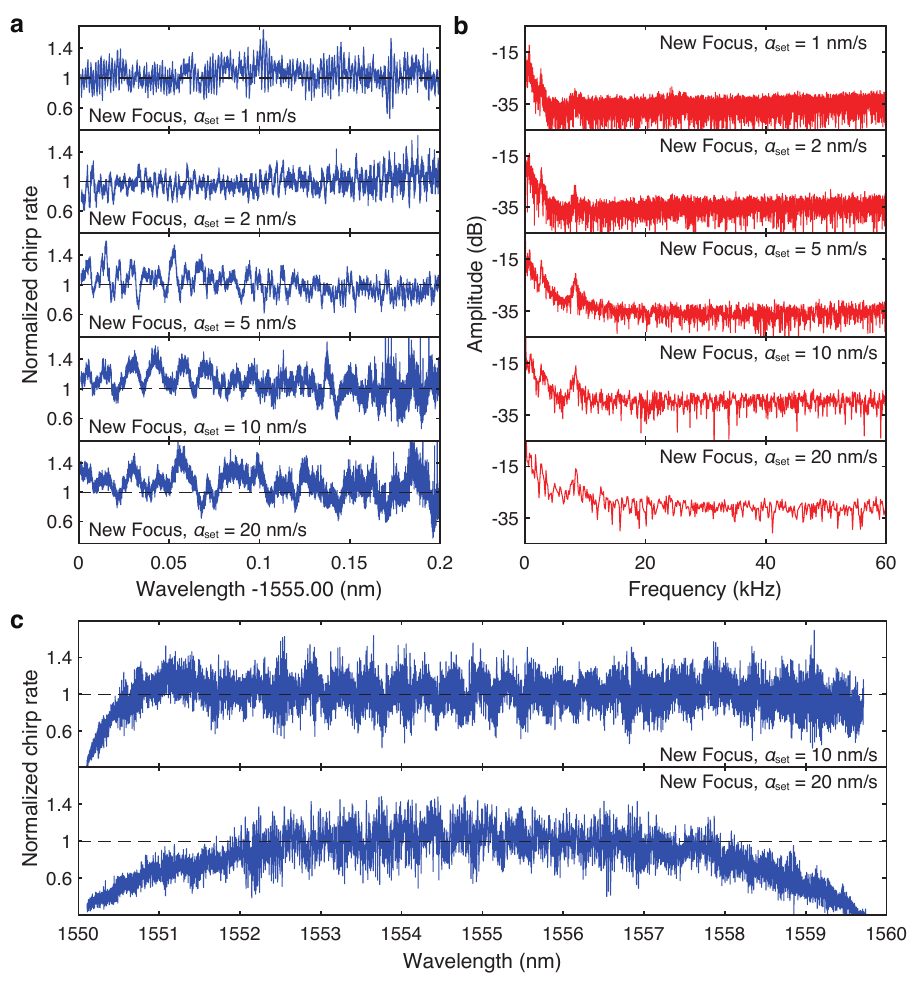}
\caption{
\textbf{Characterization of chirp dynamics of a New Focus laser in wide tuning mode}.
\textbf{a}.
Normalized chirp rate $\alpha(\lambda)/\alpha_\mathrm{set}$ with different set values $\alpha_\mathrm{set}$ over 0.2 nm wavelength range.
\textbf{b}. 
Frequency spectrum of the chirp rate $\alpha(t)/\alpha_\mathrm{set}$.
\textbf{c}. 
Normalized chirp rate $\alpha(\lambda)/\alpha_\mathrm{set}$ with different set values $\alpha_\mathrm{set}$ over 10 nm wavelength range.
}
\label{SIfig:Velocity_wide}
\end{figure*}

\begin{figure*}[t!]
\renewcommand{\figurename}{Supplementary Figure}
\centering
\includegraphics{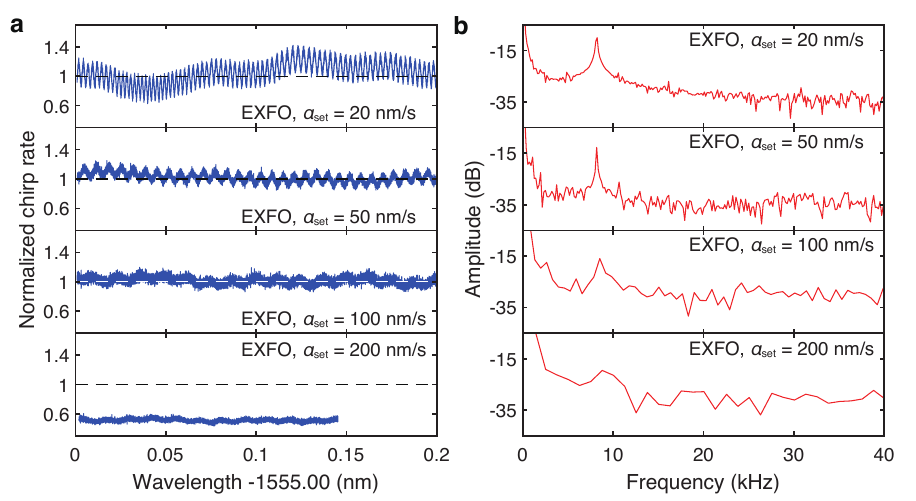}
\caption{
\textbf{Characterization of chirp dynamics of an EXFO laser in wide tuning mode}.
\textbf{a}.
Normalized chirp rate $\alpha(\lambda)/\alpha_\mathrm{set}$ with different set values $\alpha_\mathrm{set}$ over 0.2 nm wavelength range.
\textbf{b}. 
Frequency spectrum of the chirp rate $\alpha(t)/\alpha_\mathrm{set}$.
}
\label{SIfig:EXFO_wide}
\end{figure*}

\vspace{0.5cm}
For wide tuning mode, we have characterized totally 10 lasers.
These lasers include three Toptica CTL lasers, four Santec lasers, two New Focus lasers, and one EXFO laser. 
The laser chirp dynamics is investigated within 1550 to 1560 nm bandwidth, and $\alpha_\mathrm{set}$ is set to 1 nm/s, 2 nm/s, 5 nm/s, 10 nm/s, 50 nm/s, 100 nm/s, and 200 nm/s.
A summary of different lasers' chirp dynamics with different $\alpha_\mathrm{set}$ is shown in Table \ref{tab:wide}. 
As well as the laser tuning dynamics performance of the Toptica lasers shown in Figure 3 in the main text, we also present the results for other brands lasers in Fig. \ref{SIfig:Santec_wide}, Fig. \ref{SIfig:Velocity_wide}, and Fig.\ref{SIfig:EXFO_wide}.

For Santec lasers, when $\alpha_\mathrm{set}$ is relatively small, the actual chirp rate has main jitter frequencies of $2.06$ kHz, $16.04$ kHz, $32.08$ kHz, $64.10$ kHz, $96.15$ kHz and $128.17$ kHz.
These jitters become less and less pronounced as $\alpha_\mathrm{set}$ increases.
Taking into account that the random jitter of the chirp rate is slightly greater at $200$ nm/s, RMSE is larger, the optimum $\alpha_\mathrm{set}$ for the Santec laser is $100$ nm/s.
In Fig. \ref{SIfig:Santec_wide}c, we show the normalised chirp rate over the full $10$ nm chirp range for $\alpha_\mathrm{set} = 100$ nm/s.
The Santec laser performed well throughout the set chirp range, with no acceleration or deceleration in between.

As shown in Fig. \ref{SIfig:Velocity_wide}, the New Focus laser behaves essentially the same at different $\alpha_\mathrm{set}$.
The main jitter frequencies are $0.68$ kHz, $2.62$ kHz, and $8.43$ kHz.
And the performance over the full chirp range shows that the laser is not stable at the maximum chirp rate of $20$ nm/s.
When the chirp rate is 10 nm/s, the laser reaches stability at $1551.5$ nm and keeps at a chirp rate larger than the set value.

For the EXFO laser, the jitter frequency is enormous at 8.21 kHz, while no other obvious jitter frequencies are observed, as shown in Fig. \ref{SIfig:EXFO_wide}.
When $\alpha_\mathrm{set} = 200$ nm/s, the laser did not finish the chirp rate acceleration after the first 5 nm chirp range.
After changing the chirp start wavelength from 1550 nm to 1540 nm, the laser can complete its acceleration before entering the test area and shows an RMSE of 5.1\% and $\delta_\mathrm{L} = 0.2\%$.
Overall, the EXFO laser has an optima $\alpha_\mathrm{set}$ of 100 nm/s.
\begin{table*}[h!]
\centering
\caption{Laser chirp dynamics of stepper based wide tuning mode}
\label{tab:wide}
\begin{tabular}{|c|c|c|c|c|c|c|}
\hline
Laser brand & Model & Tuning mode & Chirp range & $\alpha_\mathrm{set}$ & RMSE & $\delta_\mathrm{L}$ \\
\hline
Toptica & CTL 1550 &  Wide, single & 10 nm & 1 nm/s & 16.9\% (2.4\%) & 0.2\% (0.0\%)\\
Toptica & CTL 1550 &  Wide, single & 10 nm & 2 nm/s & 7.5\% (1.1\%) & 0.2\% (0.0\%)\\
Toptica & CTL 1550 &  Wide, single & 10 nm & 5 nm/s & 26.2\% (3.4\%) & 0.5\% (0.4\%)\\
Toptica & CTL 1550 &  Wide, single & 10 nm & 10 nm/s & 14.7\% (3.3\%) & 0.3\% (0.2\%)\\
\hline
Santec & TSL-570-A & Wide, single & 10 nm & 5 nm/s & 14.5\% (6.0\%) & 0.5\% (0.3\%)\\
Santec & TSL-570-A & Wide, single & 10 nm & 10 nm/s & 11.3\% (3.9\%) & 0.7\% (0.3\%)\\
Santec & TSL-570-A & Wide, single & 10 nm & 20 nm/s & 14.0\% (6.3\%) & 1.4\% (0.8\%)\\
Santec & TSL-570-A & Wide, single & 10 nm & 50 nm/s & 6.9\% (1.8\%) & 0.4\% (0.3\%)\\
Santec & TSL-570-A & Wide, single & 10 nm & 100 nm/s & 6.8\% (1.6\%) & 0.4\% (0.2\%)\\
Santec & TSL-570-A & Wide, single & 10 nm & 200 nm/s & 8.9\% (1.6\%) & 0.3\% (0.2\%)\\
\hline
New Focus & TLB-6700 &  Wide, single & 10 nm & 1 nm/s & 13.9\% (1.1\%) & 0.4\% (0.0\%)\\
New Focus & TLB-6700 &  Wide, single & 10 nm & 2 nm/s & 12.9\% (1.9\%) & 0.3\% (0.2\%)\\
New Focus & TLB-6700 &  Wide, single & 10 nm & 5 nm/s & 13.6\% (1.4\%) & 0.8\% (0.4\%)\\
New Focus & TLB-6700 &  Wide, single & 10 nm & 10 nm/s & 14.8\% (4.0\%) & 0.9\% (0.6\%)\\
New Focus & TLB-6700 &  Wide, single & 10 nm & 20 nm/s & 17.1\% (3.4\%) & 0.6\% (0.4\%)\\
\hline
EXFO & T500S &  Wide, single & 10 nm & 20 nm/s & 15.1\% & 1.6\% \\
EXFO & T500S &  Wide, single & 10 nm & 50 nm/s & 6.5\% & 0.5\% \\
EXFO & T500S &  Wide, single & 10 nm & 100 nm/s & 4.8\% & 0.2\% \\
EXFO & T500S &  Wide, single & 10 nm & 200 nm/s & 48.9\% & 0.2\% \\
\hline
\end{tabular}
\end{table*}

\vspace{1cm}
\section{Laser chirp dynamics for piezo-driven fine tuning mode}
\begin{figure*}[h!]
\renewcommand{\figurename}{Supplementary Figure}
\centering
\includegraphics{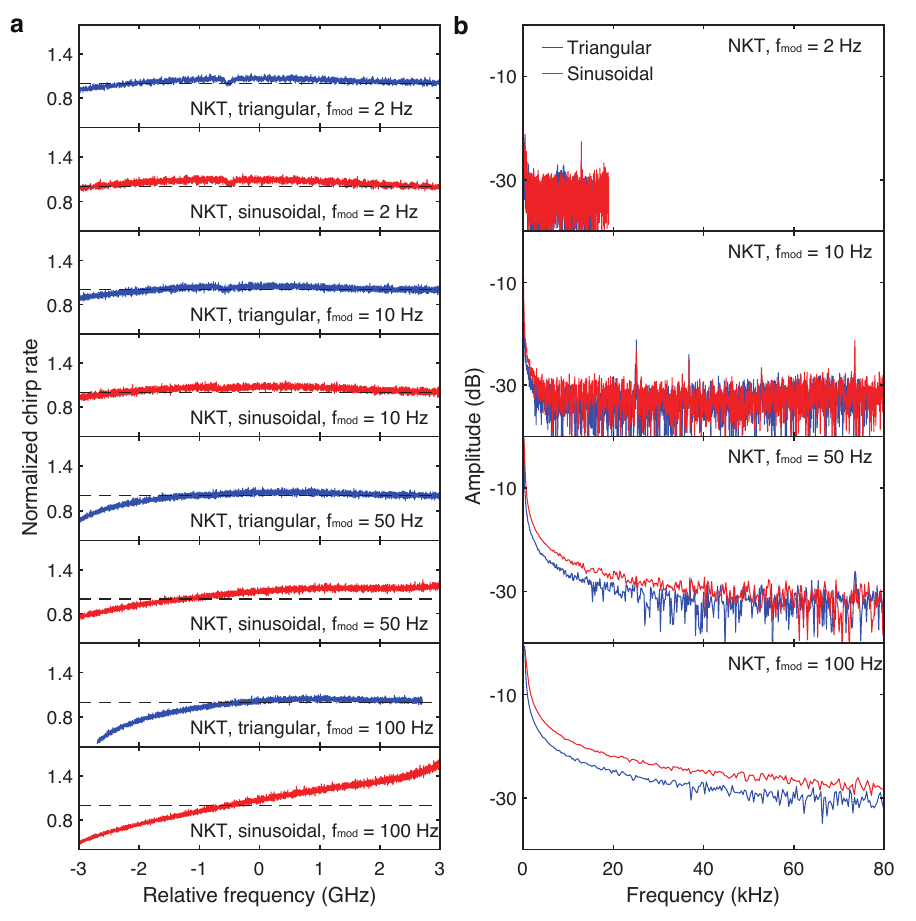}
\caption{
\textbf{Characterization of chirp dynamics of an NKT laser in fine tuning mode}.
\textbf{a}.
Normalized chirp rate $\alpha(f)/\alpha_\mathrm{set}$ with different set values $f_\text{mod}$ over 6 GHz frequency range.
The piezo of the laser is driven by a triangular (blue) or a sinusoidal (red) signal.
\textbf{b}. 
Frequency spectrum of the chirp rate $\alpha(t)/\alpha_\mathrm{set}$.
}
\label{SIfig:Nkt_piezo}
\end{figure*}

\begin{figure*}[h!]
\renewcommand{\figurename}{Supplementary Figure}
\centering
\includegraphics{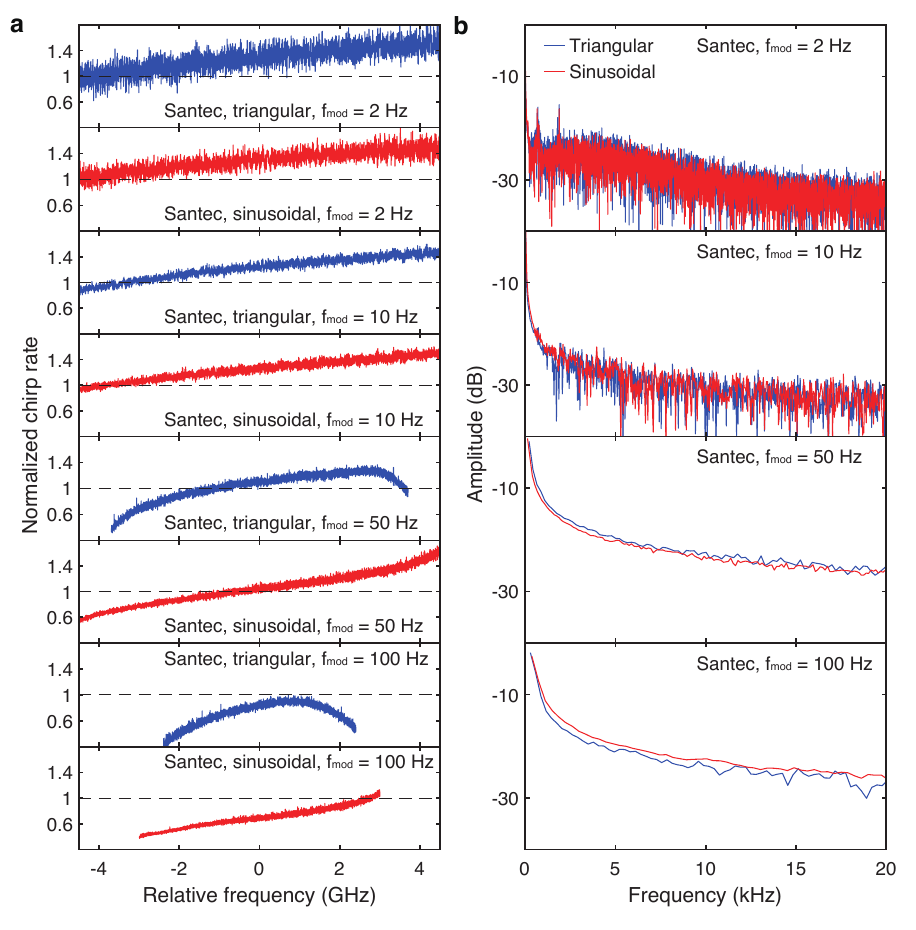}
\caption{
\textbf{Characterization of chirp dynamics of a Santec laser in fine tuning mode}.
\textbf{a}.
Normalized chirp rate $\alpha(f)/\alpha_\mathrm{set}$ with different set values $f_\text{mod}$ over 9 GHz frequency range.
The piezo of the laser is driven by a triangular (blue) or a sinusoidal (red) signal.
\textbf{b}. 
Frequency spectrum of the chirp rate $\alpha(t)/\alpha_\mathrm{set}$.
}
\label{SIfig:Santec_piezo}
\end{figure*}

\begin{figure*}[h!]
\renewcommand{\figurename}{Supplementary Figure}
\centering
\includegraphics{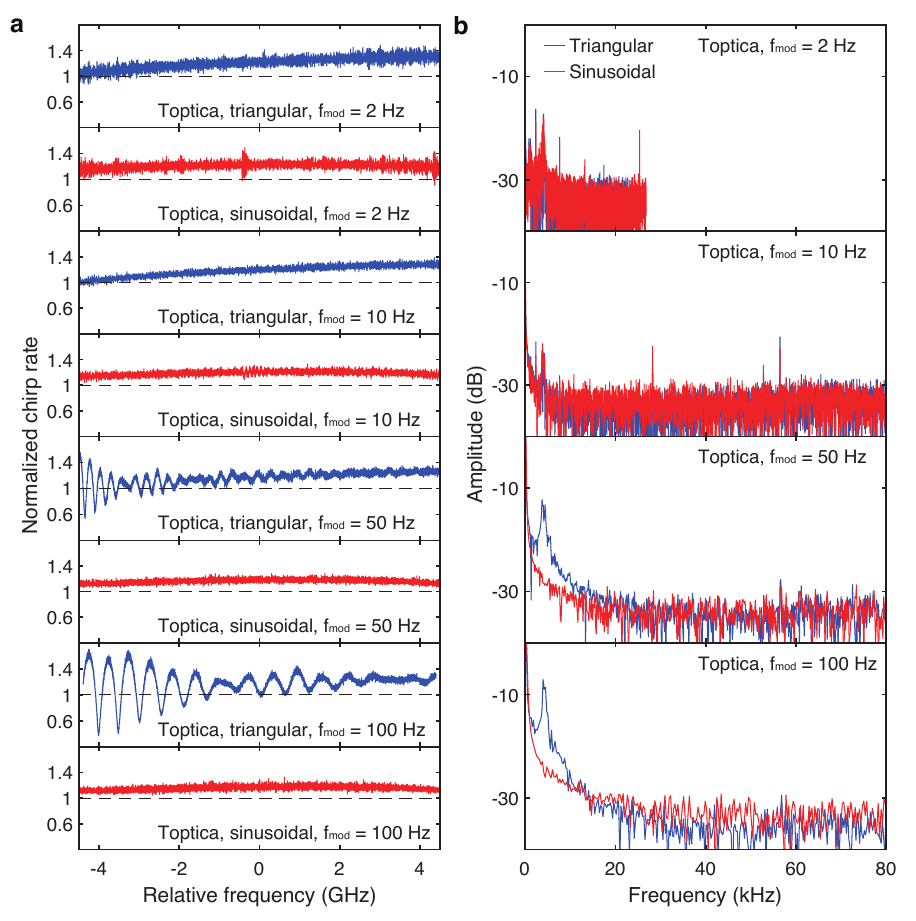}
\caption{
\textbf{Characterization of chirp dynamics of a Toptica laser in fine tuning mode}.
\textbf{a}.
Normalized chirp rate $\alpha(f)/\alpha_\mathrm{set}$ with different set values $f_\text{mod}$ over 9 GHz frequency range.
The piezo of the laser is driven by a triangular (blue) or a sinusoidal (red) signal.
\textbf{b}. 
Frequency spectrum of the chirp rate $\alpha(t)/\alpha_\mathrm{set}$.
}
\label{SIfig:Toptica_piezo}
\end{figure*}

\begin{figure*}[h!]
\renewcommand{\figurename}{Supplementary Figure}
\centering
\includegraphics{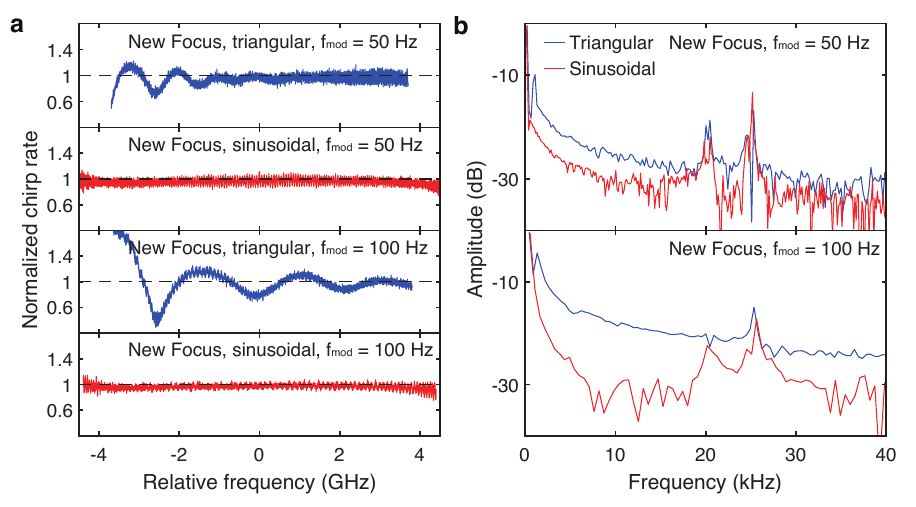}
\caption{
\textbf{Characterization of chirp dynamics of a New Focus laser in fine tuning mode}.
\textbf{a}.
Normalized chirp rate $\alpha(f)/\alpha_\mathrm{set}$ with different set values $f_\text{mod}$ over 9 GHz frequency range.
The piezo of the laser is driven by a triangular (blue) or a sinusoidal (red) signal.
\textbf{b}. 
Frequency spectrum of the chirp rate $\alpha(t)/\alpha_\mathrm{set}$.
}
\label{SIfig:Velocity_piezo}
\end{figure*}

\vspace{0.5cm}
In addition to stepper based wide tuning mode, we also characterised laser tuning dynamics of piezo-driven fine tuning-mode.
The measurement includes three Toptica lasers, four Santec lasers, two New Focus lasers, and two NKT fiber lasers.
These lasers are frequency-modulated at the central frequency of $193.3$ THz and with an excursion range of $10$ GHz ($8$ GHz for NKT lasers). 
The drive voltage signal to the piezo is either triangular or sinusoidal.
Their modulation frequencies $f_\text{mod}$ are set to $2$ Hz, $10$ Hz, $50$ Hz and $100$ Hz.
A summary of different lasers' chirp dynamics with different $f_\text{mod}$ is shown in Table \ref{tab:piezo}.
The typical results for these lasers are shown in Fig. \ref{SIfig:Nkt_piezo}, Fig. \ref{SIfig:Santec_piezo}, Fig. \ref{SIfig:Toptica_piezo} and Fig. \ref{SIfig:Velocity_piezo}.

NKT lasers behave the same with both triangular and sinusoidal drive signals.
The laser has significant jitter frequencies of $25.09$ kHz and $73.49$ kHz when $f_\text{mod}=10$ Hz.
Jitter at these frequencies disappears at higher $f_\text{mod}$. 
However, as $f_\text{mod}$ increases, so does the chirp rate drift.
Another manifestation of the NKT laser is that its actual chirp range does not reach the set value when $f_\text{mod}$ is relative high.
The optimum $f_\text{mod}$ for the NKT laser is $2$ Hz, which actually works well at different modulation frequencies.

The behavior of Santec lasers are similar to NKT ones.
The laser has significant jitter frequencies of $0.71$ kHz and $1.88$ kHz at $f_\text{mod} = 2$ Hz and these jitters disappear at higher $f_\text{mod}$.
Besides, as $f_\text{mod}$ increases, the actual chirp range of Santec lasers decreases.
Over all, Santec lasers work best when driven by a $50$ Hz sinusoidal wave.

\begin{figure*}[h!]
\renewcommand{\figurename}{Supplementary Figure}
\centering
\includegraphics{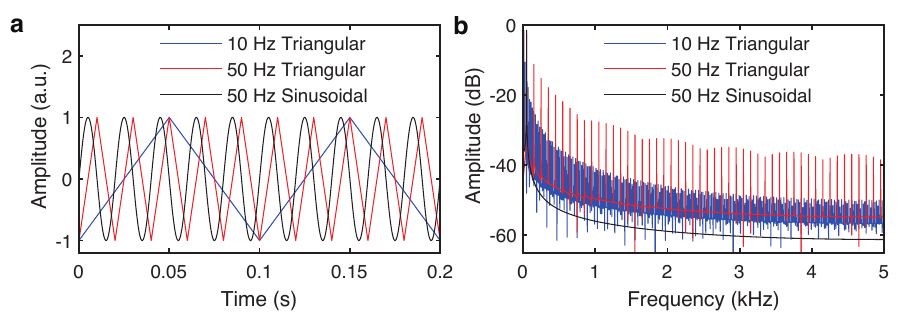}
\caption{
\textbf{a}.
Different drive signals.
\textbf{b}. 
The Fourier transform of different drive signals.
}
\label{SIfig:FFT}
\end{figure*}

The results for Toptica lasers are shown in Fig. \ref{SIfig:Toptica_piezo}.
For the triangular drive signal of $50$ Hz and $100$ Hz, an obvious chirping rate oscillation can be observed. 
While for the situation a sinusoidal signal is input, the oscillation is not evident. 
From the frequency domain, as shown in Fig. \ref{SIfig:Toptica_piezo}b, the oscillation frequency can be directly obtained as $3.84$ kHz, 
which is likely the resonant frequency of the fundamental mechanical mode excited by high-order components of the high $f_\text{mod}$ triangular input signal.
To make this easier to understand, three different drive signals are shown in Fig. \ref{SIfig:FFT}a and their Fourier transforms are shown in Fig. \ref{SIfig:FFT}b.
It can be seen that triangular waves contain many high frequency components.
The higher the drive frequency, the higher the amplitude around the $3.84$ kHz frequency of the triangular waveform.
Thus, the triangular wave drive signal tends to excite the resonance of the laser cavity.
Besides, at a modulation frequency of $10$ Hz, higher order mechanical modes with frequencies of $28.22$ kHz and $56.43$ kHz can also be observed.
Similarly, as the chirp rate increases, these obvious high-frequency resonances disappear.
Overall, Toptica lasers should be driven sinusoidally and their optimum modulation frequency is 50 Hz.

Fig. \ref{SIfig:Velocity_piezo} shows the results of New Focus lasers.
New Focus lasers exist mode-hopping when $f_\text{mod}$ equals to 10 or 20 Hz, so we failed to trace the laser's instantaneous frequency.
Besides, the laser's fundamental mechanical mode of $1.10$ kHz will be excited by a triangular driven signal.
The laser's optimum $f_\text{mod}$ is 100 Hz.

\begin{table*}[h!]
\centering
\caption{Laser chirp dynamics of piezo-driven fine tuning mode}
\label{tab:piezo}
\begin{tabular}{|c|c|c|c|c|c|c|}
\hline
Laser brand & Model & Tuning mode & $f_\text{mod}$ (Hz) & Chirp range & RMSE & $\delta_\mathrm{L}$ \\
\hline
Toptica & CTL 1550 & Piezo, triangular & 2 & 10 GHz & 17.8\% (4.5\%) &  0.8\% (0.0\%)\\
Toptica & CTL 1550 & Piezo, sinusoidal & 2 & 10 GHz  & 17.8\% (4.0\%) & -\\
Toptica & CTL 1550 & Piezo, triangular& 10 & 10 GHz & 15.8\% (3.8\%) &  0.8\% (0.0\%)\\
Toptica & CTL 1550 & Piezo, sinusoidal & 10 & 10 GHz  & 16.0\% (3.4\%) & -\\
Toptica & CTL 1550 & Piezo, triangular& 50 & 10 GHz & 16.3\% (3.4\%) &  0.9\% (0.0\%)\\
Toptica & CTL 1550 & Piezo, sinusoidal & 50 & 10 GHz  & 15.8\% (3.2\%) & -\\
Toptica & CTL 1550 & Piezo, triangular& 100 & 10 GHz & 21.5\% (5.7\%) &  0.9\% (0.0\%)\\
Toptica & CTL 1550 & Piezo, sinusoidal & 100 & 10 GHz  & 17.8\% (3.0\%) & -\\
\hline
Santec & TSL-570-A & Piezo, triangular & 2 & 10 GHz & 34.9\% (3.4\%) &  1.8\% (0.1\%)\\
Santec & TSL-570-A & Piezo, sinusoidal & 2 & 10 GHz  & 35.3\% (7.5\%) & -\\
Santec & TSL-570-A & Piezo, triangular & 10 & 10 GHz & 27.1\% (1.7\%) &  1.8\% (0.1\%)\\
Santec & TSL-570-A & Piezo, sinusoidal & 10 & 10 GHz  & 29.3\% (2.9\%) & -\\
Santec & TSL-570-A & Piezo, triangular & 50 & 10 GHz & 21.3\% (0.7\%) &  2.8\% (0.1\%)\\
Santec & TSL-570-A & Piezo, sinusoidal & 50 & 10 GHz  & 26.1\% (1.4\%) & -\\
Santec & TSL-570-A & Piezo, triangular & 100 & 10 GHz & 34.1\% (1.0\%) &  3.3\% (0.3\%)\\
Santec & TSL-570-A & Piezo, sinusoidal & 100 & 10 GHz  & 34.4\% (1.5\%) & -\\
\hline
New Focus & TLB-6700 & Piezo, triangular & 50 & 10 GHz & 18.2\% (11.8\%) &  0.4\% (0.1\%)\\
New Focus & TLB-6700 & Piezo, sinusoidal & 50 & 10 GHz  & 12.6\% (8.4\%) & -\\
New Focus & TLB-6700 & Piezo, triangular & 100 & 10 GHz & 33.2\% (7.3\%) &  1.3\% (0.1\%)\\
New Focus & TLB-6700 & Piezo, sinusoidal & 100 & 10 GHz  & 8.7\% (3.8\%) & -\\
\hline
NKT & E15 & Piezo, triangular & 2 & 8 GHz & 5.9\% (1.3\%) &  0.5\% (0.3\%)\\
NKT & E15 & Piezo, sinusoidal & 2 & 8 GHz  & 7.1\% (0.5\%) & -\\
NKT & E15 & Piezo, triangular & 10 & 8 GHz & 5.9\% (1.7\%) &  0.6\% (0.3\%)\\
NKT & E15 & Piezo, sinusoidal & 10 & 8 GHz  & 7.8\% (1.7\%) & -\\
NKT & E15 & Piezo, triangular & 50 & 8 GHz & 11.2\% (3.7\%) &  1.2\% (0.5\%)\\
NKT & E15 & Piezo, sinusoidal & 50 & 8 GHz  & 18.1\% (2.3\%) & -\\
NKT & E15 & Piezo, triangular & 100 & 8 GHz & 18.4\% (4.2\%) &  2.2\% (0.5\%)\\
NKT & E15 & Piezo, sinusoidal & 100 & 8 GHz  & 34.8\% (3.0\%) & -\\
\hline
\end{tabular}
\end{table*}
\clearpage

\vspace{1cm}
\section{Laser tuning dynamics for LiDAR demonstration}
\begin{figure*}[h!]
\renewcommand{\figurename}{Supplementary Figure}
\centering
\includegraphics{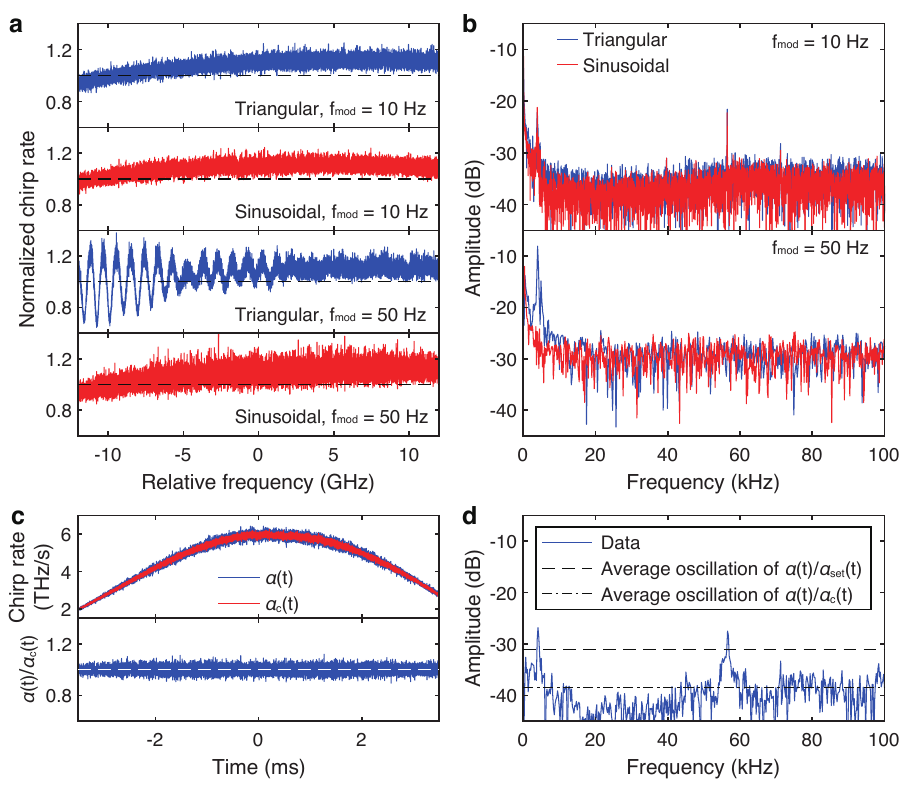}
\caption{
\textbf{Laser characterization for LiDAR demonstration}.
\textbf{a}.
Normalized chirp rate $\alpha(f)/\alpha_\mathrm{set}$ with different set values $f_\text{mod}$ over 24 GHz frequency range.
The piezo of the laser is driven by a triangular (blue) or a sinusoidal (red) signal.
\textbf{b}. 
Frequency spectrum of the chirp rate $\alpha(t)/\alpha_\mathrm{set}$.
\textbf{c}. 
Upper panel shows the chirp rate curve $\alpha(t)$ (blue) and pre-linearization curve $\alpha_\mathrm{c}(t)$ (red).
Lower panel shows the linearized chirp rate curve $\alpha(t)/\alpha_\mathrm{c}(t)$.
\textbf{d}. 
Frequency spectrum of the linearized chirp rate in lower panel of panel c.
The dot-and-dash line shows its averaged amplitude below 100 kHz.
The dashed line shows the averaged amplitude of normalized chirp rate $\alpha(t)/\alpha_\mathrm{set}(t)$ 's frequency spectrum below 100 kHz.
}
\label{SIfig:Toptica_lidar}
\end{figure*}

\vspace{0.5cm}
In FMCW LiDAR, the fine tuning mode is commonly used because of its low duty cycle.
Here we chose a Toptica laser for the LiDAR demonstration because it has the largest chirp range, 35 GHz, out of our 4 types of lasers.
We make the laser chirp with the central frequency of $193.3$ THz and the chirp range of $35$ GHz.
Some typical results are shown in Fig. \ref{SIfig:Toptica_lidar}.
Still, we can observe the fundamental mode of $3.8$ kHz excited by $50$ Hz triangular driven signal.
Besides, higher-order mechanical modes can also be observed from the upper panel of the Fig. \ref{SIfig:Toptica_lidar}b, i.e., with a frequency of 56.4 kHz. 
As $f_\text{mod}$ increases, the noise floor of the frequency-domain chirp rate increases but some obvious high-frequency resonances disappear, as shown in the lower panel of Fig. \ref{SIfig:Toptica_lidar}b.
Overall, this laser performs best when driven by a 50 Hz sinusoidal signal.

Next, we characterized the effect of our pre-linearization method.
We first measure the laser's chirp rate for three times.
The averaged result is used as the pre-linearization chirp rate $\alpha_\mathrm{c}(t)$, as shown in the upper panel of Fig. \ref{SIfig:Toptica_lidar}c.
The pre-obtained $\alpha_\mathrm{c}(t)$ can be used to perform chirp rate linearization. 
To verify the effectiveness of linearization, we characterized the laser again as the real-time measurement $\alpha(t)$.
Their consistency is compared by $\alpha(t)/\alpha_\mathrm{c}(t)$ as shown in the lower panel of Fig. \ref{SIfig:Toptica_lidar}c.
Clearly, the chirp rate drift is eliminated.
Its Fourier transform spectrum is shown in Fig.\ref{SIfig:Toptica_lidar}d.
The averaged amplitude of it below 100 kHz is shown by a dot-and-dash line.
Compared with the averaged amplitude of $\alpha(t)/\alpha_\mathrm{set}(t)$'s frequency spectrum, shown by the dashed line, the oscillation level has been attenuated by $7.4$ dB. 
However, oscillations of $3.8$ kHz and $56.4$ kHz still exist, indicating the random nature of the oscillation generated by the cavity resonances. 
This fact implies that the avoidance of the cavity resonance is crucial to reduce the phase noise of an offline calibrated chirping laser.
Thus in our case, a 50-Hz sinusoidal drive signal is utilized for LiDAR demonstration. 

\vspace{1cm}
\section{Set up for LiDAR demonstration and precision test}
\begin{figure*}[h!]
\renewcommand{\figurename}{Supplementary Figure}
\centering
\includegraphics{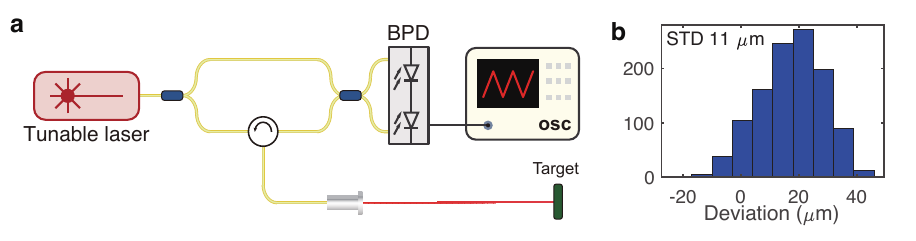}
\caption{Setup for LiDAR demonstration and precision test}
\textbf{a}.
Experimental setup. 
BPD, balanced photodetector. 
OSC, oscilloscope. 
\textbf{b}.
Histogram of deviation of range measurement.
The long-term stability of pre-linearization is verified by a precision test.
A mirror is fixed at a distance of $53.361$ mm and measured 1128 times every 3 s.
The standard deviation of the measured distance is $11$ $\mu$m. 
\label{SIfig:Lidar-precision}
\end{figure*}

\vspace{0.5cm}
With the pre-obtained $\alpha_\mathrm{c}(t)$, the LiDAR demonstration can be proceeded without any extra linearization unit.
As shown in Fig. \ref{SIfig:Lidar-precision}a, the frequency modulated laser is split into two paths.
In one path, the laser passes through a collimator, enters into free space, is reflected by the target, and is re-collected.
The reflected laser is then combined with the other reference path, and the beat signal between the two paths is recorded by a balanced photodetector.

To test the long-term stability of pre-linearization, a fixed mirror is used as the target at a distance of $53.361$ mm and measured 1128 times every 3 s.
Their relative distance histogram is shown in Fig. \ref{SIfig:Lidar-precision}b with a standard deviation of $11$ $\mu$m. 

